\documentclass[twocolumn,preprintnumbers,amsmath,amssymb]{revtex4}
\usepackage{amsmath}
\usepackage[dvips]{graphicx}
\usepackage{graphics}
\usepackage{amssymb}

\newcommand{\ppt }{P_{\alpha}^{(+)}(t)}
\newcommand{\dea }{\Delta\epsilon^{\alpha}}
\newcommand{\dva }{\Delta V^{\alpha}}
\newcommand{\del }{\Delta\epsilon^{L}}
\newcommand{\der }{\Delta\epsilon^{R}}

\begin{document}

\title{A First-Principles Method for Open Electronic Systems}
\author{Xiao Zheng}
\author{Fan Wang}
\author{GuanHua Chen}
\email{ghc@everest.hku.hk} \affiliation{Department of Chemistry,
The University of Hong Kong, Hong Kong, China}

\date{\today}


\begin{abstract}
We prove that the electron density function of a real physical
system can be uniquely determined by its values on any finite
subsystem. This establishes the existence of a rigorous
density-functional theory for any open electronic system. By
introducing a new density functional for dissipative interactions
between the reduced system and its environment, we subsequently
develop a time-dependent density-functional theory which depends in
principle only on the electron density of the reduced system. In the
steady-state limit, the conventional first-principles nonequilibrium
Green's function formulation for the current is recovered. A
practical scheme is proposed for the new density functional: the
wide-band limit approximation, which is applied to simulate the
transient current through a model molecular device.
\end{abstract}

\maketitle

\section {Introduction \label{introduction}}

Density-functional theory (DFT) has been widely used as a research
tool in condensed matter physics, chemistry, materials science, and
nanoscience. The Hohenberg-Kohn theorem~\cite{hk} lays the
foundation of DFT. The Kohn-Sham formalism~\cite{ks} provides a
practical solution to calculate the ground state properties of
electronic systems. Runge and Gross extended DFT further to
calculate the time-dependent properties and hence the excited state
properties of any electronic systems~\cite{tddft}. The accuracy of
DFT or time-dependent DFT (TDDFT) is determined by the
exchange-correlation (XC) functional. If the exact XC functional
were known, the Kohn-Sham formalism would have provided the exact
ground state properties, and the Runge-Gross extension, TDDFT, would
have yielded the exact time-dependent and excited states properties.
Despite their wide range of applications, DFT and TDDFT have been
mostly limited to isolated systems.

Many systems of current research interest are open systems. A
molecular electronic device is one such system. DFT-based
simulations have been carried out on such devices~\cite{prllang,
prlheurich, jcpluo, langprb, prbguo, jacsywt, jacsgoddard,
transiesta, jcpratner}. These simulations focus on steady-state
currents under bias voltages. Two types of approaches have been
adopted. One is the Lippmann-Schwinger formalism by Lang and
coworkers~\cite{langprb}. The other is the first-principles
nonequilibrium Green's function (NEGF) technique~\cite{prbguo,
jacsywt, jacsgoddard, transiesta, jcpratner}. In both approaches the
Kohn-Sham Fock operator is taken as the effective single-electron
model Hamiltonian, and the transmission coefficients are calculated
within the noninteracting electron model. The investigated systems
are not in their ground states, and applying ground state DFT
formalism for such systems is only an approximation~\cite{cpdatta}.
DFT formalisms adapted for current-carrying systems have also been
proposed recently, such as Kosov's Kohn-Sham equations with direct
current~\cite{jcpkosov} and Burke \emph{et al.}'s Kohn-Sham master
equation including dissipation to phonons~\cite{prlburke}. However,
practical implementation of these formalisms requires the electron
density function of the entire system. In this paper, we present a
rigorous DFT formalism for open electronic systems, and use it to
simulate the steady and transient currents through molecular
electronic devices. The first-principles formalism depends only on
the electron density function of the reduced system.

This paper is organized as follows. In Sec.~\ref{formalism} we
propose a TDDFT formalism for open electronic systems based on the
equation of motion (EOM) for reduced single-electron density matrix.
In Sec.~\ref{analyticity} we prove the theorem that the electron
density function of any finite subsystem can determine uniquely all
properties of a connected real physical system. By utilizing this
theorem we introduce in Sec.~\ref{dissipativefunc} a dissipation
functional for the electron density of the subsystem, and thus
establish a rigorous and efficient first-principles formalism for
steady and transient dynamics of open electronic systems. An
wide-band limit (WBL) approximation scheme for the dissipation
functional is proposed for practical implementations in
Sec.~\ref{schemes}. To demonstrate the applicability of our
first-principles formalism, a TDDFT calculation is carried out to
simulate the transient current through a model molecular device. The
detailed procedures and results are described in
Sec.~\ref{tddft-wbl}. Discussion and summary are given in
Sec.~\ref{summary}.

\section{First-principles formalism \label{formalism}}

\subsection{Reduced single-electron density matrix and
TDDFT formalism for reduced system \label{formalism_a}}

Fig.~\ref{scheme} depicts an open electronic system. Region $D$ is
the reduced system of our interests, and the electrodes $L$ and $R$
are the environment. Altogether $D$, $L$ and $R$ form the entire
system. Taking Fig.~\ref{scheme} as an example, we develop a
practical DFT formalism for the open systems. Within the TDDFT
formalism, a closed EOM has been derived for the reduced
single-electron density matrix $\sigma(t)$ of the entire
system~\cite{ldmtddft}:
\begin{equation}\label{eom4sigma0}
    i\dot{\sigma}(t) = [h(t),\sigma(t)],
\end{equation}
where $h(t)$ is the Kohn-Sham Fock matrix of the entire system, and
the square bracket on the right-hand side (RHS) denotes a
commutator. The matrix element of $\sigma$ is defined as
$\sigma_{ij}(t) = \langle a^{\dagger}_{j}(t)\,a_{i}(t)\rangle$,
where $a_{i}(t)$ and $a^{\dagger}_{j}(t)$ are the annihilation and
creation operators for atomic orbitals $i$ and $j$ at time $t$,
respectively. Fourier transformed into frequency domain while
considering linear response only, Eq.~(\ref{eom4sigma0}) leads to
the conventional Casida's equation~\cite{casida}. Expanded in the
atomic orbital basis set, the matrix representation of $\sigma$ can
be partitioned as
\begin{equation}\label{matrixsigma}
    \sigma = \left[\begin{array}{lll}
    \sigma_{L}  & \sigma_{LD} & \sigma_{LR} \\
    \sigma_{DL} & \sigma_{D}  & \sigma_{DR} \\
    \sigma_{RL} & \sigma_{RD} & \sigma_{R}
    \end{array}\right],
\end{equation}
where $\sigma_{L}$, $\sigma_{R}$ and $\sigma_{D}$ represent the
diagonal blocks corresponding to the left lead $L$, the right lead
$R$ and the device region $D$, respectively; $\sigma_{LD}$ is the
off-diagonal block between $L$ and $D$; and $\sigma_{RD}$,
$\sigma_{LR}$, $\sigma_{DL}$, $\sigma_{DR}$ and $\sigma_{RL}$ are
similarly defined. The Kohn-Sham Fock matrix $h$ can be
partitioned in the same way with $\sigma$ replaced by $h$ in
Eq.~(\ref{matrixsigma}). Thus, the EOM for $\sigma_{D}$ can be
written as
\begin{eqnarray}\label{eom4sigmad0}
    i\dot{\sigma}_{D} &=& [h_{D},\sigma_{D}] + \sum_{\alpha=L,R}
    \left(h_{D\alpha}\sigma_{\alpha D}-\sigma_{D\alpha}
    h_{\alpha D}\right) \nonumber \\
    &=& [h_{D},\sigma_{D}] - i\sum_{\alpha=L,R}Q_{\alpha},
\end{eqnarray}
where $Q_{L}$ ($Q_{R}$) is the dissipative term due to $L$ ($R$).
With the reduced system $D$ and the leads $L/R$ spanned
respectively by atomic orbitals $\{l\}$ and single-electron states
$\{k_{\alpha}\}$, Eq.~(\ref{eom4sigmad0}) is equivalent to:
\begin{eqnarray}\label{eom4sigmad1}
    i\dot{\sigma}_{nm} &=& \sum_{l\in
    D}\,(h_{nl}\sigma_{lm}-\sigma_{nl}h_{lm}) - i\sum_{\alpha=L,R}
    Q_{\alpha,nm}, \label{eom4sigmad2} \\
    Q_{\alpha,nm} &=& i\sum_{k_{\alpha}\in\alpha}\big(h_{nk_{\alpha}}
    \sigma_{k_{\alpha}m}-\sigma_{nk_{\alpha}}
    h_{k_{\alpha}m}\big),\label{qterm0}
\end{eqnarray}
where $m$ and $n$ correspond to the atomic orbitals in region $D$;
$k_{\alpha}$ corresponds to an electronic state in the electrode
$\alpha$ ($\alpha = L$ or $R$). $h_{nk_{\alpha}}$ is the coupling
matrix element between the atomic orbital $n$ and the electronic
state $k_{\alpha}$. The current through the interfaces $S_L$ or
$S_R$ (see Fig.~\ref{scheme}) can be evaluated as follows,
\begin{eqnarray}\label{jcurrent}
    J_{\alpha}(t) &=& -\int_{\alpha}d\mathbf{r}\,\frac{\partial}
    {\partial t}\rho(\mathbf{r},t) \nonumber \\
    &=& -\sum_{k_{\alpha}\in\alpha}\frac{d}
    {dt}\,\sigma_{k_{\alpha}k_{\alpha}}\!(t) \nonumber \\
    &=& i\sum_{l\in D}\sum_{k_{\alpha}\in\alpha}\big(
    h_{k_{\alpha}l}\,\sigma_{lk_{\alpha}} -
    \sigma_{k_{\alpha}l}\,h_{lk_{\alpha}}\big) \nonumber \\
    &=& -\sum_{l\in D}Q_{\alpha,ll}
    = -\mbox{tr}\big[Q_{\alpha}(t)\big],
\end{eqnarray}
\emph{i.e.}, the trace of $Q_{\alpha}$.

\subsection{Solution for steady-state current \label{sscurrent}}

Based on the Keldysh formalism~\cite{keldysh} and the analytic
continuation rules of Langreth~\cite{langreth}, $Q_{\alpha,nm}(t)$
can be calculated by the NEGF formulation as described in
Reference~\cite{prb94win} (see Appendix~\ref{derivee4qt})
\begin{eqnarray}\label{e4qt}
    Q_{\alpha,nm}(t)&=&-\sum_{l\in D}\int_{-\infty}^{\infty}d\tau
    \Big[\,G^{<}_{nl}(t,\tau)\Sigma^{a}_{\alpha,lm}(\tau,t)
    +\nonumber\\
    &&\,G^{r}_{nl}(t,\tau)\Sigma^{<}_{\alpha,lm}(\tau,t)
    -\Sigma^{<}_{\alpha,nl}(t,\tau)G^{a}_{lm}(\tau,t)\nonumber\\
    &&-\,\Sigma^{r}_{\alpha,nl}(t,\tau)G^{<}_{lm}(\tau,t)\Big],
\end{eqnarray}
where $G^{r}$, $G^{a}$ and $G^{<}$ are the retarded, advanced and
lesser Green's function for the reduced system $D$, respectively,
and $\Sigma^{r}_{\alpha}$, $\Sigma^{a}_{\alpha}$ and
$\Sigma^{<}_{\alpha}$ are the retarded, advanced and lesser
self-energies due to the lead $\alpha$ ($L$ or $R$), respectively.
Combining Eqs.~(\ref{jcurrent}) and (\ref{e4qt}), we obtain
\begin{eqnarray}\label{jqttddft}
    J_{\alpha}(t) &=& 2\Re\,\bigg\{\int_{-\infty}^{\infty}\!d\tau\,
    \,\mbox{tr}\Big[G^{<}_{D}(t,\tau)\Sigma^{a}_{\alpha}(\tau,t)+
    \nonumber\\
    &&G^{r}_{D}(t,\tau)\Sigma^{<}_{\alpha}(\tau,t)\Big]\bigg\}.
\end{eqnarray}
The same expression was derived by Stefanucci and Almbladh within
the framework of TDDFT~\cite{qttddft}.

It is important to point out that
Eqs.~(\ref{eom4sigma0})$-$(\ref{qterm0}) follow the partition-free
scheme proposed by Cini~\cite{cini}, while Eq.~(\ref{e4qt}) was
derived if one follows the partitioned scheme developed by Caroli
\emph{et al.}~\cite{caroli}. In the above derivation we assume
that the equivalence of the two schemes, which is satisfied if the
two self-energies behave asymptotically as follows~\cite{qttddft},
\begin{eqnarray}\label{self-e-t-tau}
    \lim_{t\rightarrow \infty}\Sigma^{r}_{\alpha}(t,t') =
    \lim_{t\rightarrow \infty}\Sigma^{a}_{\alpha}(t',t) = 0.
\end{eqnarray}

As $t,\tau\rightarrow +\infty$, $\Gamma^{k_{\alpha}}_{nm}(t,\tau)
=h_{nk_ {\alpha}}(t)\,h_{k_{\alpha}m}(\tau)$ becomes
asymptotically time-independent. The Green's functions for the
reduced system $D$ rely simply on the difference of the two
time-variables~\cite{qttddft}, and can thus be expressed as
\begin{eqnarray}\label{expandedglra}
   G^{<}_{nm}(t,\tau) &=& \sum_{p,q\in D}\int_{-\infty}
   ^{\infty}dt_{1}\int_{-\infty}^{\infty}dt_{2}\,G^{r}_{np}
   (t,t_{1})\nonumber \\
   &&\times\,\Sigma^{<}_{pq}(t_{1},t_{2})\,G^{a}_{qm}(t_{2},\tau)\nonumber\\
   &=& i\sum_{p,q\in D}\sum_{\alpha=L,R}\sum_{l_{\alpha}\in\alpha}f^
   {\alpha}_{l}\nonumber \\
   &&\times\left[\int_{-\infty}^{\infty}dt_{1}\textnormal{e}^
   {-i\epsilon_{l}^{\alpha}t_{1}}G^{r}_{np}(t-t_{1})\right]\Gamma
   ^{l_{\alpha}}_{pq}\nonumber\\
   &&\times\left[\int_{-\infty}^{\infty}dt_{2}\textnormal{e}
   ^{i\epsilon_{l}^{\alpha}t_{2}}G^{a}_{qm}(t_{2}-\tau)\right]
   \nonumber\\
   &=& i\sum_{p,q\in D}\sum_{\alpha=L,R}\sum_{l_{\alpha}\in\alpha}
   f^{\alpha}_{l}\textnormal{e}^{-i\epsilon^{\alpha}_{l}(t-\tau)}
   \nonumber \\
   && \times\,G^{r}_{np}(\epsilon^{\alpha}_{l})\,\Gamma^{l_{\alpha}}_{pq}\,
   G^{a}_{qm}(\epsilon^{\alpha}_{l}),\label{g01}\\
   G^{r,a}_{nm}(\epsilon) &=& \left[\epsilon I-h-\Sigma^{r,a}(\epsilon)
   \right]^{-1}_{nm}, \label{g02}\\
   \Sigma^{r,a}_{nm}(\epsilon) &=&
   \sum_{\alpha=L,R}\sum_{l\in\alpha}\,
   \Gamma^{l_{\alpha}}_{nm}\left(\epsilon-\epsilon_{l}^{\alpha}\pm
   i\Delta\right)^{-1}, \label{g03}
\end{eqnarray}
where $I$ is the identity matrix. The steady-state current can
thus be explicitly expressed by combining
Eqs.~(\ref{g01})$-$(\ref{g03}),
\begin{eqnarray}\label{jss}
    J_{L}(\infty) &=& -J_{R}(\infty) \nonumber \\
    &=& -\sum_{n\in D}Q_{L,nn}(\infty) \nonumber\\
    &=& 2\pi\,\Bigg\{\sum_{k\in L}f^{L}_{k}\sum_{l\in R}\Delta
    (\epsilon^{R}_{l}-\epsilon^{L}_{k})\nonumber \\
    && \times\,\mbox{tr}\Big[G^{r}_
    {D}(\epsilon^{L}_{k})\,\Gamma^{l_{R}}\,G^{a}_{D}(\epsilon^{L}
    _{k})\,\Gamma^{k_{L}}\Big] \nonumber \\
    &&-\sum_{l\in R}f^{R}_{l}\sum_{k\in L}
    \delta(\epsilon^{L}_{k}-\epsilon^{R}_{l})\nonumber \\
    &&\times\,\mbox{tr}\Big[
    G^{r}_{D}(\epsilon^{R}_{l})\,\Gamma^{l_{R}}\,G^{a}_{D}
    (\epsilon^{R}_{l})\,\Gamma^{k_{L}}\Big]\Bigg\} \nonumber\\
    &=&\int\left[f^{L}(\epsilon)-f^{R}(\epsilon)\right]T(\epsilon)
    \,d\epsilon,\\
    T(\epsilon)&=&2\pi\,\eta_{L}\eta_{R}\,\mbox{tr}
    \Big[G^{r}_{D}(\epsilon)\Gamma^{R}(\epsilon)G^{a}_{D}
    (\epsilon)\Gamma^{L}(\epsilon)\Big].\label{tofe}
\end{eqnarray}
Here $T(\epsilon)$ is the transmission coefficient,
$f^{\alpha}(\epsilon)$ is the Fermi distribution function, and
$\eta_{\alpha}(\epsilon)= \sum_{k\in\alpha}\delta(\epsilon -\epsilon
^{\alpha}_{k})$ is the density of states (DOS) for the lead $\alpha$
($L$ or $R$). Eq.~(\ref{jss}) is exactly the Landauer
formula~\cite{bookdatta, landauer} in the DFT-NEGF
formalism~\cite{prbguo, jacsywt}. The difficulty in solving
Eq.~(\ref{eom4sigmad2}) is to calculate $Q_{\alpha,nm}$. Employing
the Keldysh NEGF formalism, the evaluation of $Q_{\alpha,nm}$
involves the calculation of two-time Green's functions and
self-energies as those appearing in Eq.~(\ref{e4qt}), which makes
the simulation of any real molecular device computationally
impractical. An alternative approach must be developed.

\section{Holographic electron density theorem for time-dependent systems
\label{analyticity}}

As early as in 1981, Riess and M\"{u}nch~\cite{riess} discovered the
holographic electron density theorem which states that any nonzero
volume piece of the ground state electron density determines the
electron density of a molecular system. This is based on that the
electron density functions of atomic and molecular eigenfunctions
are real analytic away from nuclei. In 1999 Mezey extended the
holographic electron density theorem~\cite{mezey}. And in 2004
Fournais~\emph{et al.} proved again the real analyticity of the
electron density functions of any atomic or molecular
eigenstates~\cite{analyticity}. Therefore, for a time-independent
real physical system made of atoms and molecules, its electron
density function is real analytic (except at nuclei) when the system
is in its ground state, any of its excited eigenstates, or any state
which is a linear combination of finite number of its eigenstates;
and the ground state electron density on any finite subsystem
determines completely the electronic properties of the entire
system.

\begin{figure}
\includegraphics[scale=0.45]{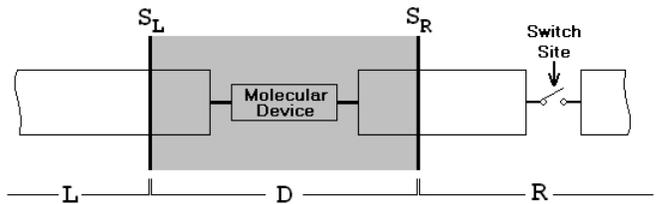}
\caption{\label{scheme} Schematic representation of the experimental
setup for quantum transport through a molecular device.}
\end{figure}

As for time-dependent systems, the issue is less clear. Although it
seems intuitive that the electron density function of any
time-dependent real physical system is real analytic (except for
isolated points in space-time), it turns out quite difficult to
prove the analyticity rigorously. Fortunately we are able to
establish a one-to-one correspondence between the electron density
function of any finite subsystem and the external potential field
which is real analytic in both $t$-space and $\mathbf{r}$-space, and
thus circumvent the difficulty concerning the analyticity of
time-dependent electron density function. For time-dependent real
physical systems, we have the following theorem:

{\it Theorem:} If the electron density function of a real physical
system at $t_0$, $\rho(\mathbf{r},t_0)$, is real analytic in
$\mathbf{r}$-space, the corresponding wave function is $\Phi(t_0)$,
and the system is subjected to a real analytic (in both $t$-space
and $\mathbf{r}$-space) external potential field $v(\mathbf{r},t)$,
the time-dependent electron density function on any finite subspace
$D$, $\rho_D(\mathbf{r},t)$, has a one-to-one correspondence with
$v(\mathbf{r},t)$ and determines uniquely all electronic properties
of the entire time-dependent system.

{\it Proof:} Let $v(\mathbf{r},t)$ and $v'(\mathbf{r},t)$ be two
real analytic potentials in both $t$-space and $\mathbf{r}$-space
which differ by more than a constant at any time $t\geqslant t_0$,
and their corresponding electron density functions are
$\rho(\mathbf{r},t)$ and $\rho'(\mathbf{r},t)$, respectively.
Therefore, there exists a minimal nonnegative integer $k$ such that
the $k$-th order derivative differentiates these two potentials at
$t_0$:
\begin{equation}
   \left.\frac{\partial^k}{\partial t^k}\left[v(\mathbf{r},t) -
   v'(\mathbf{r},t)\right]\right\vert_{t=t_0} \neq \mbox{const}.
   \label{v-vp}
\end{equation}
Following exactly the Eqs.~(3)-(6) of Ref.~\cite{tddft}, we have
\begin{eqnarray}
   \left.\frac{\partial^{k+2}}{\partial t^{k+2}}\left[
   \rho(\mathbf{r},t)-\rho'(\mathbf{r},t)\right]\right\vert_{t=t_0}
   &=& -\nabla\cdot u(\mathbf{r}), \label{rho-rhop}
\end{eqnarray}
where
\begin{eqnarray}
   u(\mathbf{r}) &=& \rho(\mathbf{r},t_0)\,\nabla\!\left\{
   \left.\frac{\partial^k}{\partial t^k}\left[v(\mathbf{r},t) -
   v'(\mathbf{r},t)\right]\right\vert_{t=t_0}\right\}.
   \label{uofr}
\end{eqnarray}
Due to the analyticity of $\rho(\mathbf{r},t_0)$, $v(\mathbf{r},t)$
and $v'(\mathbf{r},t)$, $\nabla\cdot u(\mathbf{r})$ is also real
analytic in $\mathbf{r}$-space. It has been proven in
Ref.~\cite{tddft} that it is \emph{impossible} to have $\nabla\cdot
u(\mathbf{r})=0$ on the entire $\mathbf{r}$-space. Therefore it is
also impossible that $\nabla\cdot u(\mathbf{r})=0$ everywhere in $D$
because of analytical continuation of $\nabla\cdot u(\mathbf{r})$.
Note that $\rho_D(\mathbf{r},t) = \rho(\mathbf{r},t)$ for
$\mathbf{r}\in D$. We have thus
\begin{eqnarray}
   \left.\frac{\partial^{k+2}}{\partial t^{k+2}}\left[
   \rho_{D}(\mathbf{r},t)-\rho_{D}'(\mathbf{r},t)\right]\right
   \vert_{t=t_0}&\neq& 0 \label{rhow-rho'w}
\end{eqnarray}
for $\mathbf{r}\in D$. This confirms the existence of a one-to-one
correspondence between $v(\mathbf{r},t)$ and $\rho_D(\mathbf{r},t)$.
$\rho_{D}(\mathbf{r},t)$ thus determines uniquely all electronic
properties of the entire system. This completes the proof of the
{\it Theorem}.

Note that if $\Phi(t_0)$ is the ground state, any excited
eigenstate, or any state as a linear combination of finite number of
eigenstates of a time-independent Hamiltonian, the prerequisite
condition in {\it Theorem} that the electron density function
$\rho(\mathbf{r},t_0)$ be real analytic is automatically satisfied,
as proven in Ref.~\cite{analyticity}. As long as the electron
density function at $t=t_0$, $\rho(\mathbf{r},t_0)$, is real
analytic, it is guaranteed that $\rho_D(\mathbf{r},t)$ of the
subsystem $D$ determines all physical properties of the entire
system at any time $t$ if the external potential $v(\mathbf{r},t)$
is real analytic.

According to the above \emph{Theorem}, the electron density function
of any subsystem determines all the electronic properties of the
entire time-dependent physical system. This proves in principle the
existence of a rigorous DFT-type formalism for open electronic
systems. All one needs to know is the electron density of the
reduced system.

\section{Dissipative density functional \label{dissipativefunc}}

According to the holographic electron density theorem of
time-dependent physical systems, all physical quantities are
explicit or implicit functionals of the electron density in the
reduced system $D$, $\rho_{D}(\mathbf{r},t)$. $Q_{\alpha}$ of
Eq.~(\ref{eom4sigmad0}) is thus also a universal functional of
$\rho_{D}(\mathbf{r},t)$. Therefore, Eq.~(\ref{eom4sigmad1}) can
be recast into a formally closed form,
\begin{equation}\label{e4rdm}
  \!\!i\dot{\sigma_{D}}= \Big[h_{D}[\mathbf{r},t;\rho_{D}
    (\mathbf{r},t)],\sigma_{D}\Big]-i\!\!\sum_{\alpha=L,R}
    \!\!Q_{\alpha}[\mathbf{r},t;\rho_{D}(\mathbf{r},t)].
\end{equation}
Neglecting the second term on the RHS of Eq.~(\ref{e4rdm}) leads to
the conventional TDDFT formulation in terms of reduced
single-electron density matrix~\cite{ldmtddft} for the isolated
reduced system. The second term describes the dissipative processes
between $D$ and $L$ or $R$. Besides the XC functional, an additional
universal density functional, the dissipation functional
$Q_{\alpha}[\mathbf{r},t; \rho_{D}(\mathbf{r},t)]$, is introduced to
account for the dissipative interaction between the reduced system
and its environment. Eq.~(\ref{e4rdm}) is the TDDFT EOM for open
electronic systems. It would thus be much more efficient integrating
Eq.~(\ref{e4rdm}) than solving Eqs.~(\ref{eom4sigmad1}) and
(\ref{e4qt}), if $Q_{\alpha}[\mathbf{r},t;\rho_{D}(\mathbf{r},t)]$
or its approximation is known. We therefore have a practical and
potentially rigorous formalism for any open electronic systems.
Burke \emph{et al.} extended TDDFT to include electronic systems
interacting with phonon baths~\cite{prlburke}, they proved the
existence of a one-to-one correspondence between $v(\mathbf{r},t)$
and $\rho(\mathbf{r},t)$ under the condition that the dissipative
interactions (denoted by a superoperator $\mathcal{C}$ in
Ref.~\cite{prlburke}) between electrons and phonons are fixed. In
our case since the electrons can move in and out the reduced system,
the number of the electrons in the reduced system is not conserved.
In addition, the dissipative interactions can be determined in
principle by the electron density of the reduced system. We do not
need to stipulate that the dissipative interactions with the
environment are fixed as Burke \emph{et al.}. And the only
information we need is the electron density of the reduced system.
In the frozen DFT approach~\cite{warshel} an additional XC
functional term was introduced to account for the XC interaction
between the system and the environment. This additional term is
included in $h_D[\mathbf{r},t;\rho_{D}(\mathbf{r},t)]$ of
Eq.~(\ref{e4rdm}).

\begin{figure}
\includegraphics[scale=0.4]{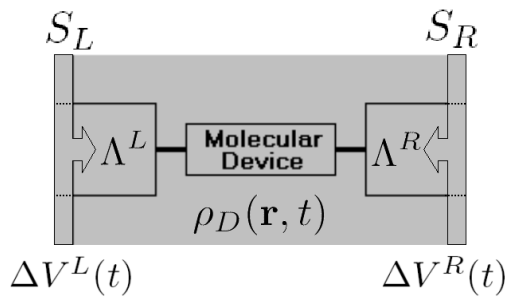}
\caption{\label{device} Schematic representation of the reduced
system $D$ within the WBL scheme for dissipation functional
$Q_{\alpha}$. }
\end{figure}

Given $Q_{\alpha}[\rho_D(\mathbf{r},t)]$ how do we solve the
EOM~(\ref{e4rdm}) in practice? Again take the molecular device
shown in Fig.~\ref{scheme} as an example. We focus on the reduced
system $D$ as depicted in Fig.~\ref{device}, and integrate the
EOM~(\ref{e4rdm}) directly by satisfying the boundary conditions
at $S_{L}$ and $S_{R}$. $\Delta V^L(t)$ and $\Delta V^R(t)$ are
the bias voltages applied on $L$ and $R$, respectively, and serve
as the boundary conditions at $S_L$ and $S_R$, respectively. At
$t\rightarrow -\infty$, $\Delta V^L = \Delta V^R = 0$, and near $t
= 0$ $\Delta V^L(t)$ and $\Delta V^R(t)$ are turned on
adiabatically. We need thus integrate Eq.~(\ref{e4rdm}) together
with a Poisson equation for Coulomb potential inside the device
region $D$. And the Poisson equation is subjected to the boundary
condition determined by the potentials at $S_{L}$ and $S_{R}$.

\section{Wide-band limit approximation for dissipation
functional $\mathbf{Q}_{\mathbf{\alpha}}$ and its test on a model
system \label{schemes}}

An explicit form for the dissipation functional $Q_{\alpha}$ is
required for practical implementation of Eq.~(\ref{e4rdm}).
Admittedly $Q_{\alpha}[\mathbf{r},t; \rho_{D}(\mathbf{r},t)]$ is an
extremely complex functional and difficult to evaluate. As various
approximated expressions have been adopted for the DFT XC functional
in practical implementations, the same strategy can be applied to
the dissipation functional $Q_{\alpha}$.

One such scheme is the wide-band limit (WBL)
approximation~\cite{prb94win} which involves the following
assumptions for the leads: (i) their band-widths are assumed to be
infinitely large, (ii) their line-widths,
$\Lambda^{\alpha}_{k}(t,\tau)$, defined by the DOS at $S_{L}$ or
$S_{R}$ times the coupling strength between $D$ and $L$ or $R$,
\emph{i.e.}, $\Lambda^{\alpha}_{k}(t,\tau)=\pi\,\eta_{\alpha}
(\epsilon^{\alpha}_{k})\,\Gamma^{k_{\alpha}}(t,\tau)$, are treated
as energy independent, \emph{i.e.}, $\Lambda^{\alpha}_{k}(t,\tau)
\approx \Lambda^{\alpha}(t,\tau) \approx \Lambda^\alpha$, and (iii)
the level shifts of $L$ or $R$ are taken as a constant for all
energy levels, \emph{i.e.}, $\Delta\epsilon^{\alpha}_{k}(t) \approx
\Delta\epsilon^{\alpha}(t) = -\Delta V^{\alpha}(t)$, where $\Delta
V^{\alpha}(t)$ are bias voltages applied on $L$ or $R$ at time $t$.
The detailed derivations for the WBL scheme can be found in
Appendix~\ref{EOM4WBL} and the explicit form for $Q^{WBL}_{\alpha}$
is given here,
\begin{equation}\label{efinal}
    Q^{WBL}_\alpha(t) = K^{\alpha}(t)
    +\left\{\Lambda^{\alpha},\sigma_{D}(t)\right\},
\end{equation}
Here $K^{\alpha}(t)$ is fully expanded as follows,
\begin{eqnarray}\label{k-wbl-1}
   K^{\alpha}(t) &=& -\frac{2i}{\pi}\,\bigg\{ \,U^{\alpha}(t)
   \int_{-\infty}^{\mu^{0}}\frac{d\epsilon\,
   \mbox{e}^{i\epsilon t}}{\epsilon-h_{D}(0)+i\,\Lambda} \nonumber
   \\
   && + \int_{-\infty}^{\mu^{0}}\left[ I - U^{\alpha}(t)\,
   \mbox{e}^{i\epsilon t} \right] \nonumber \\
   &&  \times \frac{d\epsilon }{\epsilon-h_{D}(t)+i\,\Lambda
   + \Delta\epsilon^{\alpha}(t)}
   \bigg\}\Lambda^{\alpha} + H.C.
\end{eqnarray}
where
\begin{eqnarray}\label{u-alpha}
   U^{\alpha}(t) &=& \mbox{e}^{-i\int_0^t \left[h_D(\tau) -i\Lambda
   -\dea(\tau)\right]d\tau}.
\end{eqnarray}
From Eqs.~(\ref{efinal})-(\ref{u-alpha}) it is clear that the
dissipation functional $Q_{\alpha}$ within WBL scheme depends
explicitly on $\dea(t)$, $\sigma_{D}(t)$, $h_{D}(t)$ and
$\Lambda^{\alpha}$. Note that $\dea(t) = -\Delta V^{\alpha}(t)$
and $\dva(t)$ is a functional of $\rho_D(\mathbf{r},t)$,
\emph{i.e.}, $\dva(t) \equiv \dva\left[ \rho_D(\mathbf{r},t), t
\right]$; $h_D(t) \equiv h_D\left[ \sigma_D(t), t \right]$; since
$\eta_{\alpha}(\epsilon)$ is the DOS at $S_{\alpha}$ ($\alpha = L$
or $R$), and $\Gamma^{\alpha}$ is the coupling strength between
the surface states at $S_{\alpha}$ and the bulk states of $D$,
$\Lambda^{\alpha}$ is thus a functional of $\rho_D(\mathbf{r},t)$,
\emph{i.e.}, $\Lambda^{\alpha} \equiv \Lambda^{\alpha}\left[
\rho_D(\mathbf{r},t), t\right]$. We hence conclude that in
practice $Q^{WBL}_{\alpha}$ is a functional of
$\rho_D(\mathbf{r},t)$, \emph{i.e.},
\begin{eqnarray}\label{q-of-rho-d}
  Q^{WBL}_{\alpha}(t) &\equiv& Q^{WBL}_{\alpha}\Big[
  \sigma_D\!\left[\rho_D\right],
  \,h_D[\sigma_D\!\left[\rho_D\right],t], \nonumber \\
  && \Lambda^{\alpha}[\rho_D,t], \,\dva[\rho_D,t], \,t \Big].
\end{eqnarray}

\begin{figure}
\includegraphics[scale=0.35]{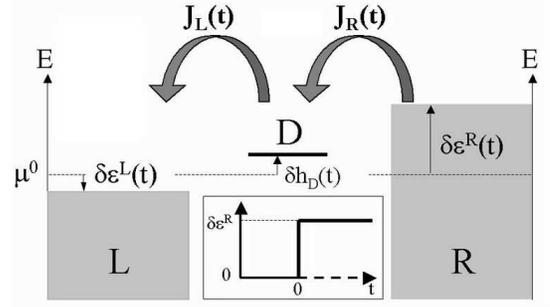}
\caption{\label{onesite} Model system for the test of the WBL
dissipation functional where a single site spans the device region
$D$. Transient currents through leads $L$ and $R$, $J_L(t)$ and
$J_R(t)$, are simulated. The inset shows the time-dependent level
shift of lead $R$. }
\end{figure}

\begin{figure}
\includegraphics[scale=0.7]{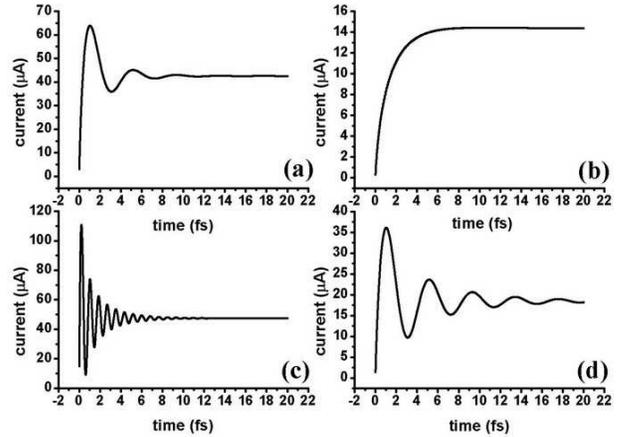}
\caption{\label{wbl-awbl} The calculated transient current through
$S_R$ within the WBL scheme. We set $\mu^0 = h_{D}(0) = 0$ for the
ground state; and $\Delta\epsilon^L(t) = 0$ and
$\Delta\epsilon^R(t) = \Delta\epsilon^R\, (1 - \mbox{e}^{-t / a})$
after switch-on. The above panels show different cases where (a)
$\Delta\epsilon^R = 2$ eV, $\Lambda^L = \Lambda^R = 0.1$ eV; (b)
$\Delta\epsilon^R = 0.2$ eV, $\Lambda^L = \Lambda^R = 0.1$ eV; (c)
$\Delta\epsilon^R = 10$ eV, $\Lambda^L = \Lambda^R = 0.1$ eV; and
(d) $\Delta\epsilon^R = 2$ eV, $\Lambda^L = \Lambda^R = 0.04$ eV,
respectively. }
\end{figure}

The WBL dissipation functional $Q^{WBL}_{\alpha}$ is then tested
by calculations on a model system which has previously been
investigated by Maciejko, Wang and Guo~\cite{pulse2c}. In this
model system the device region $D$ consists of a single site
spanned by only one atomic orbital (see Fig.~\ref{onesite}). Exact
transient current driven by a step voltage pulse has been obtained
from NEGF simulations~\cite{pulse2c}, and the authors concluded
that the WBL approximation yields reasonable results provided that
the band-widths of the leads are five times or larger than the
coupling strength between $D$ and $L$ or $R$. The computational
details are as follows. The entire system ($L$ + $R$ + $D$) is
initially in its ground state with the chemical potential $\mu^0$.
External bias voltages are switched on from the time $t = 0$,
which results in transient current flows through the leads $L$ and
$R$. $\delta h_D(t) \equiv h_D(t) - h_D(0)$, $\del(t)$ and
$\der(t)$ are the level shifts of $D$, $L$ and $R$ at time $t$,
respectively. In our works we take $\delta h_D(t) = \frac{1}{2}
\left[\del(t) + \der(t)\right]$, $\Delta\epsilon^L(t) = 0$, and
$\Delta\epsilon^R(t) = \Delta\epsilon^R\, (1 - \mbox{e}^{-t /
a})$, where $a$ is a positive constant. The real analytic level
shift $\Delta \epsilon^R(t)$ resembles perfectly a step pulse as
$a\rightarrow 0^{+}$. The calculation results are demonstrated in
Fig.~\ref{wbl-awbl}. We choose exactly the same parameter set as
that adopted for Fig.~2 in Ref.~\cite{pulse2c}, and the resulting
transient current, represented by Fig.~\ref{wbl-awbl}(a),
excellently reproduces the WBL result in Ref.~\cite{pulse2c},
although the numerical procedures employed are distinctively
different. The comparison confirms evidently the accuracy of our
formalism. From Fig.~\ref{wbl-awbl}(a)-(c) it is observed that
with the same line-widths $\Lambda^{\alpha}$, a larger level shift
$\Delta\epsilon^R$ results in a more fluctuating current, whereas
by comparing (a) and (d) we see that under the same
$\Delta\epsilon^R$, the current decays more rapidly to the steady
state value with larger $\Lambda^{\alpha}$.

Since the integration over energy in Eq.~(\ref{k-wbl-1}) can be
performed readily by transforming the integrand into diagonal
representation, $Q^{WBL}_{\alpha}$ are evaluated efficiently, which
makes the WBL scheme a practical routine for subsequent TDDFT
calculations.

\section{A TDDFT calculation of transient current
\label{tddft-wbl}}

\begin{figure}
\includegraphics[scale=0.45]{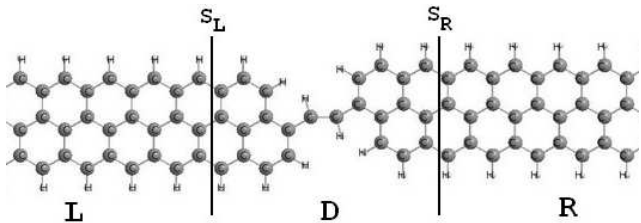}
\caption{\label{tddft-system} A two-dimensional model molecular
device is connected to left and right leads. }
\end{figure}

With the EOM~(\ref{e4rdm}) and the WBL scheme for the dissipation
functional $Q_{\alpha}$, it is now straightforward to carry out
first-principles calculations for transient dynamics of open
electronic systems. A model molecular device depicted in
Fig.~\ref{tddft-system} is taken as the open system under
investigation. The device region $D$ containing $24$ carbon and
$12$ hydrogen atoms is spanned by the 6-31 Gaussian basis set,
\emph{i.e.}, altogether $240$ basis functions for the reduced
system. The leads are quasi-one-dimensional graphene sheets with
dangling bonds saturated by hydrogen atoms, and the entire system
is on a same plane. The ground state reduced single-electron
density matrix for the reduced system, $\sigma_D(0)$, is extracted
from $\sigma(0)$ of an extended system which consists of totally
$134$ atoms, covering not only the device region $D$ but also
portions of leads $L$ and $R$. This provides the initial condition
for the EOM~(\ref{e4rdm}). The line-widths $\Lambda^{L}$ and
$\Lambda^{R}$ within the WBL scheme are obtained from the surface
Green's functions for isolated semi-infinite bulk leads $L$ and
$R$, $g^{r}_{L}(\mu^{0})$ and $g^{r}_{R}(\mu^{0})$~\cite{surfg},
respectively, and then optimized such that the RHS of the
EOM~(\ref{e4rdm}) vanishes correctly at $t = 0$. The molecular
device is switched on by a step-like voltage $\Delta V^R(t) =
-\Delta\epsilon^R(t) = \Delta V^R (1 - \mbox{e}^{-t / a})$ applied
on the right lead (see the inset of Fig.~\ref{onesite}), while
$\Delta V^L(t) = 0$, and the dynamic response of the reduced
system is obtained by solving the EOM~(\ref{e4rdm}) in time domain
within the adiabatic local density approximation
(ALDA)~\cite{casida} for the XC functional and the WBL
approximation for the dissipation functional. To save
computational resources we linearize the XC component of the
induced Kohn-Sham Fock matrix on $D$, $\delta h^{XC}_D(t)$, as
follows,
\begin{eqnarray}
   \delta h^{XC}_{ij}(t) &=& \sum_{mn\in D}V^{XC}_{ijmn}\,
   \left[\sigma_{mn}(t) - \sigma_{mn}(0)\right],\label{xcksf}\\
V^{XC}_{ijmn} &=& \int_D d\mathbf{r}\,\phi^{\ast}_m(\mathbf{r})
\phi_n(\mathbf{r})\frac{\delta v^{XC}[\rho_D(\mathbf{r},t)]}
{\delta\rho_D(\mathbf{r},t)} \nonumber \\
&& \times\,\phi^{\ast}_i(\mathbf{r}) \phi_j(\mathbf{r}),
\label{vxcijmn}
\end{eqnarray}
where $v^{XC}[\rho_D(\mathbf{r},t)]$ is the XC potential. The
Coulomb component of $\delta h_D(t)$ is constructed by solving the
Poisson equation for the device region $D$ subjected to boundary
conditions $\Delta V^{\alpha}(t)$ at every time $t$. The TDDFT
calculations are carried out with a modified version of the
TDDFT-LDM program developed by Yam, Yokojima and
Chen~\cite{ldmtddft}.

\begin{figure}
\includegraphics[scale=0.65]{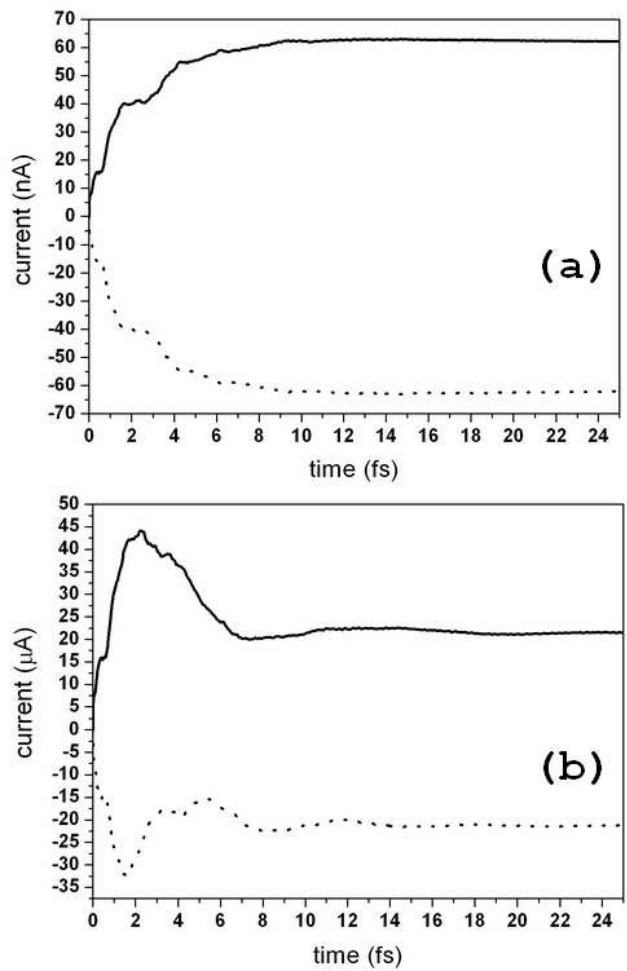}
\caption{\label{tddft-result} The solid (dotted) line represents the
transient current through $S_R$ ($S_L$) driven by a step-like
voltage $\Delta V^R(t)$ applied on the lead $R$ with the amplitude
(a) $\Delta V^R = -1$ mV, and (b) $\Delta V^R = -1$ V. }
\end{figure}

In Fig.~\ref{tddft-result}(a) and (b) we plot the transient
currents through the interfaces $S_L$ and $S_R$, $J_L(t)$ and
$J_R(t)$, for cases where the turn-on voltage $\Delta V^R = -1$ mV
and $\Delta V^R = -1$ V, respectively. The EOM~(\ref{e4rdm}) is
integrated numerically by the fourth-order Runge-Kutta
method~\cite{kutta} up to $25$ fs with the time step $0.02$ fs.
$J_L(t)$ and $J_R(t)$ depicted in Fig.~\ref{tddft-result}(a)
increase rapidly in the first $5$ fs and then approach gradually
towards their steady state values. The steady currents through the
leads $L$ and $R$ are $-62.8$ nA and $62.8$ nA, respectively, and
thus cancel each other out exactly, as they should. With a much
larger turn-on voltage $\Delta V^R$, $J_L(t)$ and $J_R(t)$ exhibit
conspicuous overshooting during the first $2$ fs, as shown in
Fig.~\ref{tddft-result}(b), and afterwards they decay slowly to
their steady state values, \emph{i.e.}, $-21.4$ $\mu$A and $21.4$
$\mu$A, respectively. From the both cases shown in
Fig.~\ref{tddft-result} diversified fluctuations are observed for
the time-dependent currents. This is due to the various
eigenvalues possessed by the non-negative definite line-widths
$\Lambda^{\alpha}$ with their magnitudes ranging from $0$ to $39$
eV, corresponding to various dissipative channels between $D$ and
$L$ or $R$. For much higher turn-on voltages the linearized form
for $\delta h^{XC}_D(t)$ (Eq.~(\ref{xcksf})) becomes inadequate,
which makes such a TDDFT calculation computationally demanding
with our present coding. From Fig.~\ref{tddft-result}, the
characteristic switch-on time for the model molecular device
depicted in Fig.~\ref{tddft-system} is estimated as about $10$ fs
for applied bias voltages as large as $1$ V.

\section{Discussion and Summary \label{summary}}

With an explicit form of the universal dissipation functional
$Q_{\alpha}$, the time evolution of an open electron system in
external fields is fully characterized by the EOM for the reduced
single-electron density matrix of the reduced system (see
Eq.~(\ref{e4rdm})). In practical calculations, we need thus focus
only on the reduced system with appropriate boundary conditions.
In conventional quantum dissipation theory (QDT)~\cite{qdt} the
key quantity is the reduced system density matrix. Whereas in
Eq.~(\ref{e4rdm}) the basic variable is the reduced
single-electron density matrix, which leads to the drastic
reduction of the degrees of freedom in numerical simulation.
Linear-scaling methods such as the localized-density-matrix
method~\cite{ldmtddft,ldm} may thus be adopted to further speed up
the solution process of Eq.~(\ref{e4rdm}). Yokojima \emph{et al.}
developed a dynamic mean-field theory for dissipative interacting
many-electron systems~\cite{yokojima1,yokojima2}. An EOM for the
reduced single-electron density matrix was derived to simulate the
excitation and nonradiative relaxation of a molecule embedded in a
thermal bath. This is in analogy to our case although our
environment is actually a fermion bath instead of a boson bath.
More importantly the number of electrons in the reduced system is
conserved in Refs.~\cite{yokojima1,yokojima2} while in our case it
is not. Therefore, Eq.~(\ref{e4rdm}) provides a rigorous and
convenient formalism to investigate the dynamic properties of open
systems. Recently Cui~\emph{et al.} proposed a TDDFT scheme for
first-principles study of non-equilibrium quantum transport based
on the complete second-order quantum dissipation theory
(CS-QDT)~\cite{csqdt-scba}, their formulation is constructed in
terms of an improved reduced density matrix approach at the
self-consistent Born approximation (SCBA) level.

It is worth mentioning that our first-principles method for open
systems applies to the same phenomena, properties or systems as
those intended by Hohenberg and Kohn~\cite{hk}, Kohn and
Sham~\cite{ks}, and Runge and Gross~\cite{tddft}, {\it i.e.},
where the exchange-correlation energy is a functional of electron
density only, $E_{XC} = E_{XC}[\rho(\mathbf{r})]$. This is true
when the interaction between the electric current and magnetic
field is negligible. However, in the presence of a strong magnetic
field, $E_{XC} =
E_{XC}[\rho(\mathbf{r}),\mathbf{j}_{p}(\mathbf{r})]$ or $E_{XC} =
E_{XC}[\rho(\mathbf{r}),\mathbf{B}(\mathbf{r})]$, where
$\mathbf{j}_p(\mathbf{r})$ is the paramagnetic current density and
$\mathbf{B}(\mathbf{r})$ is the magnetic field~\cite{magdft}. In
such a case, our first-principles formalism needs to be
generalized to include $\mathbf{j}_{p}(\mathbf{r})$ or
$\mathbf{B}(\mathbf{r})$. Of course, $\mathbf{j}_{p}(\mathbf{r})$
or $\mathbf{B}(\mathbf{r})$ should be an analytical function in
space. It is important to note that our formalism applies in principle
to Cini's scheme. Caroli's scheme is employed to derive an approximated
expression for the dissipative functional $Q_{\alpha}$.

To summarize, we have proved rigorously the existence of a
first-principles method for time-dependent open electronic systems,
and developed a formally closed TDDFT formalism by introducing a new
dissipation functional. This new functional $Q_{\alpha}$ depends
only on the electron density function of the reduced system. With an
efficient WBL scheme for $Q_{\alpha}$, we have applied the
first-principles formalism to carry out a TDDFT calculation for
transient current through a model molecular device. This work
greatly extends the realm of density-functional theory.

\begin{acknowledgements}
Authors would thank Hong Guo, Shubin Liu, Jiang-Hua Lu, Zhigang
Shuai, K. M. Tsang, Jian Wang, Arieh Warshel and Weitao Yang for
stimulating discussions. Support from the Hong Kong Research Grant
Council (HKU 7010/03P) is gratefully acknowledged.
\end{acknowledgements}

\appendix
\section{Derivation of Eq.~(\ref{e4qt}) \label{derivee4qt}}

In Keldysh formalism~\cite {keldysh}, the nonequilibrium
single-electron Green's function $G_{k_{\alpha},m}(t,t')$ is
defined by
\begin{equation}\label{contourg0}
    G_{k_{\alpha}m}(t,t') = -i\left\langle T_{C}\!\left
    \{a_{k_{\alpha}}\!(t)\,a_{m}^{\dag}(t')\right\}\right\rangle,
\end{equation}
where $T_{C}$ is the contour-ordering operator along the Keldysh
contour~\cite{keldysh}. Its lesser component, $G^{<}_
{k_{\alpha},m}(t,t')$, is defined by
\begin{equation}\label{glesser0}
    G^{<}_{k_{\alpha}m}(t,t') = i\langle\,a^{\dag}_{m}(t')\,
    a_{k_{\alpha}}(t)\rangle.
\end{equation}
Therefore $\sigma_{k_{\alpha}m}(t)$ is precisely the lesser
Green's function of identical time variables, \emph{i.e.},
$\sigma_{k_{\alpha}m}(t) = -iG^{<}_{k_{\alpha}m}(t,t')\vert_
{t'=t}$. The formal NEGF theory has exactly the same structure as
that of the time-ordered Green's function at zero temperature.
Thus, the Dyson equation for $G_{k_{\alpha}m}(t,t')$ can be
written as
\begin{equation}\label{eom4gc}
    G_{k_{\alpha}m}(t,t')=\sum_{l\in D}\int_{C}d\tau\,g_{k_{\alpha}}
    (t,\tau)\,h_{k_{\alpha}l}(\tau)\,G_{lm}(\tau,t'),
\end{equation}
where $G_{lm}(\tau,t')$ and $g_{k_{\alpha}}(t,\tau)$ are the
contour-ordered Green's functions for the reduced system $D$ and
the isolated lead $\alpha$ ($L$ or $R$), respectively. Applying
the analytic continuation rules of Langreth~\cite{langreth}, we
have
\begin{eqnarray}\label{eom4gcless}
    G^{<}_{k_{\alpha}m}(t,t')&=& \sum_{l\in D}\int_{-\infty}^{\infty}
    d\tau\,h_{k_{\alpha}l}(\tau)\Big[g^{<}_{k_{\alpha}}(t,\tau)\,
    G^{a}_{lm}(\tau,t')\nonumber\\
    && +\,\,g^{r}_{k_{\alpha}}(t,\tau)\,G^{<}_{lm}(\tau,t')\Big],
\end{eqnarray}
where $G^{a}_{lm}(\tau,t')$ and $G^{<}_{lm}(\tau,t')$ are the
advanced and lesser Green's functions for the reduced system $D$,
respectively, and $g^{r}_{k_{\alpha}}(t,\tau)$ and
$g^{<}_{k_{\alpha}}(t,\tau)$ are the retarded and lesser Green's
functions for the isolated lead $\alpha$ ($L$ or
$R$)~\cite{prb94win}, respectively. Note that
\begin{eqnarray}\label{glesser-mk}
    G^{<}_{mk_{\alpha}}(t',t)\!&=& i\langle a^{\dag}_{k_{\alpha}}
    (t)\,a_{m}(t')\rangle \nonumber \\
    &=& -\left[G^{<}_{k_{\alpha}m}(t,t')\right]^{\ast}.
\end{eqnarray}
Obviously $\sigma_{mk_{\alpha}}(t) =
-iG^{<}_{mk_{\alpha}}(t',t)\vert_ {t'=t}$. Combining
Eqs.~(\ref{eom4gcless}) and (\ref{glesser-mk}), we obtain
\begin{eqnarray}\label{glesser-mk-detail}
    G^{<}_{mk_{\alpha}}(t',t)&=&\sum_{l\in D}\int_{-\infty}^{\infty}
    d\tau\,h_{lk_{\alpha}}(\tau)\Big[g^{<}_{k_{\alpha}}(\tau,t)\,
    G^{r}_{ml}(t',\tau) + \nonumber\\
    && g^{a}_{k_{\alpha}}(\tau,t)\,G^{<}_{ml}(t',\tau)\Big]
\end{eqnarray}
by employing the following equalities:
\begin{eqnarray}\label{equalities0}
    G^{r}_{ml}(t',\tau) &=& \left[G^{a}_{lm}(\tau,t')
    \right]^{\ast}, \nonumber \\
    G^{<}_{ml}(t',\tau) &=& -\left[G^{<}_{lm}(\tau,t')
    \right]^{\ast}, \nonumber \\
    g^{a}_{k_{\alpha}}(\tau,t) &=& \left[g^{r}_{k_{\alpha}}
    (t,\tau)\right]^{\ast}, \nonumber \\
    g^{<}_{k_{\alpha}}(\tau,t) &=& -\left[g^{<}_{k_{\alpha}}
    (t,\tau)\right]^{\ast}.
\end{eqnarray}
By inserting Eqs.~(\ref{eom4gcless}) and (\ref{glesser-mk-detail})
into Eq.~(\ref{eom4sigmad1}), Eq.~(\ref{e4qt}) is recovered
straightforwardly where the self-energy terms are defined by
\begin{eqnarray}\label{selfes}
    \Sigma^{a}_{\alpha,lm}(t,\tau) &=& \sum\limits_{k_{\alpha}\in\alpha}
    h_{lk_{\alpha}}(t)\,g^{a}_{k_{\alpha}}(t,\tau)\,h_{k_{\alpha}m}(\tau),
    \nonumber \\
    \Sigma^{<}_{\alpha,nl}(t,\tau) &=& \sum\limits_{k_{\alpha}\in\alpha}
    h_{nk_{\alpha}}(t)\,g^{<}_{k_{\alpha}}(t,\tau)\,h_{k_{\alpha}l}(\tau),
    \nonumber \\
    \Sigma^{r}_{\alpha,nl}(t,\tau) &=& \sum\limits_{k_{\alpha}\in\alpha}
    h_{nk_{\alpha}}(t)\,g^{r}_{k_{\alpha}}(t,\tau)\,h_{k_{\alpha}l}(\tau).
\end{eqnarray}

\section{Wide-band limit approximation for dissipation
functional $\mathbf{Q}_{\mathbf{\alpha}}$ \label{EOM4WBL}}

Within the WBL scheme, the retarded and advanced self-energies
become local in time~\cite{prb94win},
\begin{eqnarray}\label{raselfe}
    \Sigma^{a}_{\alpha,nm}(\tau,t) &=& \sum_{k_{\alpha}\in\alpha}
    h_{nk_{\alpha}}(t)h_{k_{\alpha}m}(\tau)g^{a}_{k_{\alpha}}(\tau,t)
    \nonumber \\
    &=& \sum_{k_{\alpha}\in\alpha}h_{nk_{\alpha}}(t)h_{k_{\alpha}m}
    (\tau)\nonumber \\
    &&\times\,\left[\,i\vartheta(t-\tau)\,
    \mbox{e}^{\,i\epsilon _{k}^{\alpha}(t-\tau)}\,\mbox{e}^
    {\,i\int^{t}_{\tau}\Delta\epsilon^{\alpha}(\bar{t})\,d\bar{t}}\,\right]
    \nonumber \\
    &=& \frac{i}{\pi}\,\vartheta
    (t-\tau)\,\mbox{e}^{\,i\int^{t}_{\tau}\Delta\epsilon^{\alpha}
    (\bar{t})\,d\bar{t}} \nonumber\\
    &&\times\left\{
    \int_{-\infty}^{+\infty}\mbox{e}^{i\epsilon(t-\tau)}
    d\epsilon\right\}\Lambda^{\alpha}_{nm} \nonumber \\
    &=& i\delta(t-\tau)\Lambda^{\alpha}_{nm},\\
    \Sigma^{r}_{\alpha,nm}(\tau,t) &=& \left[\Sigma^{a}_{\alpha,mn}
    (t,\tau)\right]^{\ast} \nonumber \\
    &=& -i\delta(t-\tau)\Lambda^{\alpha}_{nm}.
\end{eqnarray}
The third equality of Eq.~(\ref{raselfe}) involves the following
approximation for the line-widths within the WBL scheme,
\begin{eqnarray}
   \Lambda^{\alpha}_{k}(t,\tau) & \equiv & \pi\,\eta_{\alpha}
   (\epsilon^{\alpha}_k)\,h_{nk_ {\alpha}}(t)\,h_{k_{\alpha}m}(\tau)
   \nonumber \\
   & \approx & \Lambda^{\alpha}(t,\tau) \approx \Lambda^\alpha.
\end{eqnarray}
Initially the entire system ($D$ + $L$ + $R$) is in its ground state
with the chemical potential $\mu^0$, from the time $t = 0$ it is
switched on by external potentials $\Delta V^{\alpha}(t)$ applied on
the leads $L$ or $R$. Hence, for $t,\tau>0$ we have
\begin{eqnarray}\label{tau-gt-zero}
    \Sigma^{<}_{\alpha,nm}(\tau,t) &=& \sum_{k_{\alpha}\in\alpha}
    h_{nk_{\alpha}}(t)h_{k_{\alpha}m}(\tau)g^{<}_{k_{\alpha}}(\tau,t)
    \nonumber \\
    &=&\sum_{k_{\alpha}\in\alpha}h_{nk_{\alpha}}(t)h_{k_{\alpha}m}(\tau)
    \nonumber\\
    &&\times\left[i\,f^{\alpha}(\epsilon^{\alpha}_{k})\,
    \mbox{e}^{\,i\epsilon _{k}^{\alpha}(t-\tau)}\,\mbox{e}^
    {\,i\int^{t}_{\tau}\Delta\epsilon^{\alpha}(\bar{t})\,d\bar{t}}
    \,\right]\nonumber\\
    &=& \frac{2i}{\pi}\,\mbox{e}^{\,i\int^{t}_{\tau}\Delta\epsilon^{\alpha}
    (\bar{t})\,d\bar{t}}\,\Lambda^{\alpha}_{nm}\nonumber\\
    &&\times\left\{\int_{-\infty}^{+\infty}f^{\alpha}(\epsilon)\,\mbox{e}^
    {\,i\epsilon(t-\tau)}d\epsilon\right\}, \label{lselfe-gt-zero}\\
    G^{r}_{nm}(t,\tau)&=&-i\vartheta(t-\tau)\sum_{l\in D}U^{(-)}_{nl}
    (t)U^{(+)}_{lm}(\tau),\label{gr4wbl-gt-zero}
\end{eqnarray}
where $\dea(t) = -\Delta V^{\alpha}(t)$ are the time-dependent level
shifts for the leads $L$ and $R$, while for $\tau<0$ and $t > 0$,
the counterparts of (\ref{lselfe-gt-zero}) and
(\ref{gr4wbl-gt-zero}) are as the following:
\begin{eqnarray}
    \Sigma^{<}_{\alpha,nm}(\tau,t)&=&\sum_{k_{\alpha}\in\alpha}
    h_{nk_{\alpha}}(t)h_{k_{\alpha}m}(\tau)\,g^{<}_{k_{\alpha}}(\tau,t)
    \nonumber \\
    &=&\sum_{k_{\alpha}\in\alpha}h_{nk_{\alpha}}(t)h_{k_{\alpha}m}(\tau)
    \nonumber \\
    &&\times\,\Big[i\,f^{\alpha}(\epsilon^{\alpha}_{k})\,\mbox{e}^{\,i
    \epsilon^{\alpha}_{k}(t-\tau)}\mbox{e}^{\,i\int^{t}_{0}\Delta
    \epsilon^{\alpha}(\bar{t})\,d\bar{t}}\,\Big] \nonumber\\
    &=& \frac{2i}{\pi}\,\mbox{e}^{\,i\int^{t}_{0}\Delta
    \epsilon^{\alpha}(\bar{t})\,d\bar{t}}
    \,\Lambda^{\alpha}_{nm} \nonumber \\
    && \times\,\left\{\int_{-\infty}^{+\infty}f^{\alpha}(\epsilon)\,\mbox{e}^
    {\,i\epsilon(t-\tau)}d\epsilon\right\}, \label{lselfe-lt-zero}
\end{eqnarray}
\begin{eqnarray}
    G^{r}_{nm}(t,\tau)&=& \sum_{l\in D}U^{(-)}_{nl}(t)\,G^{r}_{lm}
    (0,\tau)\nonumber\\
    &=&\sum_{l\in D}U^{(-)}_{nl}(t)\,G^{r,\,0}_{lm}(-\tau),
    \label{gr4wbl-lt-zero}
\end{eqnarray}
where $G^{r,\,0}_{lm}(-\tau)$ is the retarded Green's function for
the reduced system $D$ before switch-on. The propagators for the
reduced system $U^{(\pm)}(t)$ are defined as
\begin{eqnarray}\label{u+u-0}
    U^{(\pm)}(t)&=&\exp\bigg\{\pm i\int_{0}^{t}\!h_{D}(\tau)d\tau
    \pm\Lambda\,t\bigg\}.
\end{eqnarray}
where $\Lambda = \sum_{\alpha=L,R}\Lambda^{\alpha}$. By inserting
Eqs.~(\ref{raselfe})$-$(\ref{gr4wbl-lt-zero}) into Eq.~(\ref{e4qt})
the explicit form of WBL approximation for the dissipation
functional $Q_\alpha$ is obtained as
\begin{eqnarray}
    Q^{WBL}_\alpha(t) &=& K^{\alpha}(t) +
    \left\{\Lambda^{\alpha},\sigma_{D}(t)\right\}, \label{qfinal0}
\end{eqnarray}
where the curly bracket on the RHS denotes an anticommutator, and
$K^{\alpha}(t)$ is a Hermitian matrix expressed by
\begin{eqnarray}
    K^{\alpha}(t) &=& P^{\alpha}(t) + \left[P^{\alpha}(t)\right]
    ^{\dagger}, \label{k-alpha-wbl}
\end{eqnarray}
where $P^{\alpha}(t)$ involve an integration over the entire
$t$-space, which is then decomposed into positive and negative
parts, denoted by $P^{(+)}_{\alpha}(t)$ and $P^{(-)}_{\alpha}(t)$,
respectively.
\begin{eqnarray}
    P^{\alpha}(t) &=& -\int_{-\infty}^{+\infty}d\tau\,
    G^{r}_{D}(t,\tau)\Sigma^{<}_{\alpha}(\tau,t) \nonumber \\
    &=& P_{\alpha}^{(-)}(t) + P_{\alpha}^{(+)}(t).\label{p-alpha}
\end{eqnarray}
$P_{\alpha}^{(-)}(t)$ and $P_{\alpha}^{(+)}(t)$ are evaluated via
\begin{eqnarray}\label{p-alpha-minus}
    P_{\alpha}^{(-)}(t) &=& -\int_{-\infty}^{0}d\tau\,
    G^{r}_{D}(t,\tau)\Sigma^{<}_{\alpha}(\tau,t)\nonumber\\
    &=&-\frac{2i}{\pi}\,\,\mbox{exp}\left\{i\!\int_{0}^{t}\Delta
    \epsilon^{\alpha}\!(\tau)d\tau\right\}U^{(-)}(t)
    \nonumber \\
    && \times \left\{\int_{-\infty}^{\mu^{0}}\frac
    {d\epsilon\,\mbox{e}^{i\epsilon t}}{\epsilon-h_{D}(0)+i\,\Lambda}
    \right\}\Lambda^{\alpha},
\end{eqnarray}
and
\begin{eqnarray} \label{p_alpha_plus_2}
   P_{\alpha}^{(+)}(t) &=& -\frac{2}{\pi}\int^{\mu^0}_{-\infty}
   d\epsilon\,W_{\alpha}^{(-)}(\epsilon,t) \nonumber \\
   && \times \int^{t}_{0}d\tau\,
   W_{\alpha}^{(+)}(\epsilon,\tau)\,\Lambda^{\alpha},
\end{eqnarray}
respectively, where
\begin{eqnarray} \label{w_alpha_plus}
   W_{\alpha}^{\pm}(\epsilon,t) &=& \mbox{e}^{ \pm\,i
   \int^{t}_{0}d\tau \left[h_{D}(\tau) -i\Lambda -
   \Delta\epsilon^{\alpha}(\tau) - \epsilon\right]}.
\end{eqnarray}
However, the evaluations of
Eqs.~(\ref{p_alpha_plus_2})-(\ref{w_alpha_plus}) are found extremely
time-consuming since at every time $t$ one needs to propagate
$W_{\alpha}^{\pm}(\epsilon,t)$ for every individual $\epsilon$
inside the lead energy spectrum. It is thus necessary to seek for a
simpler approximate form for $P^{(+)}_{\alpha}(t)$ with satisfactory
accuracy retained. Note that Eq.~(\ref{p_alpha_plus_2}) can be
reformulated as
\begin{eqnarray}
   \ppt &=& -\frac{2}{\pi}\int_{-\infty}^{\mu^0}d\epsilon
   \int_0^t d\tau \nonumber \\
   && \times\,\mbox{e}^{-i\int_{\tau}^t \left[h_D(\bar{t})-i\Lambda
   -\dea(\bar{t})-\epsilon \right]d\bar{t}}\,\Lambda^{\alpha}.
   \label{paplus1}
\end{eqnarray}
For cases where a steady state can be ultimately reached,
$\Delta\epsilon^{\alpha}(t)$ and $h_D(t)$ become asymptotically
constant as time $t\rightarrow +\infty$, \emph{i.e.},
$\dea(t)\rightarrow \dea(\infty)$ and $h_D(t)\rightarrow
h_D(\infty)$. Therefore, the steady state $P_{\alpha}^{(+)}(\infty)$
can be approximated by substituting $\dea(\infty)$ and $h_D(\infty)$
for $\dea(t)$ and $h_D(t)$ in Eq.~(\ref{paplus1}), respectively.
\begin{eqnarray}
   P_{\alpha}^{(+)}(\infty) & \approx &
   -\frac{2}{\pi}\int_{-\infty}^{\mu^0}d\epsilon \int_0^t d\tau\ \nonumber \\
   && \times\,\mbox{e}^{-i\left[h_D(\infty) -i\Lambda
   -\Delta\epsilon^{\alpha}(\infty)-\epsilon \right](t-\tau)}
   \,\Lambda^{\alpha}\nonumber \\
   &=& -\frac{2i}{\pi}\int_{-\infty}^{\mu^0} \left\{I -
   \mbox{e}^{-i\left[h_D(\infty) -i\Lambda -\dea(\infty) - \epsilon
   \right]t} \right\} \nonumber\\
   && \times\,\frac{d\epsilon}{\epsilon - h_D(\infty) + i\Lambda + \dea(\infty)}
   \,\Lambda^{\alpha}.  \label{paplus2}
\end{eqnarray}
It is obvious from Eq.~(\ref{paplus1}) that
\begin{eqnarray}
 P_{\alpha}^{(+)}(0) = 0. \label{paplus10}
\end{eqnarray}
Thus $\ppt$ for any time $t$ between $0$ and $+\infty$ can be
approximately  expressed by adiabatically connecting Eq.~(\ref{paplus2})
with (\ref{paplus10}) as follows,
\begin{eqnarray}
   P_{\alpha}^{(+)}(t) & \approx &
   -\frac{2i}{\pi}\int_{-\infty}^{\mu^0} \left\{I -
   \mbox{e}^{-i\int_0^t \left[h_D(\tau) -i\Lambda
   -\dea(\tau) - \epsilon\right]d\tau} \right\} \nonumber\\
   && \times\,\frac{d\epsilon}{\epsilon -
   h_D(t) + i\Lambda + \dea(t)}\,\Lambda^{\alpha}.
   \label{paplus3}
\end{eqnarray}
Both Eqs.~(\ref{paplus1}) and (\ref{paplus3}) lead to the correct
$P^{\alpha}(\infty)$ for steady states,
\begin{eqnarray}
   P^{\alpha}(\infty) &=& -\frac{2i}{\pi}\!\int_{-\infty}^{\mu^0}
   \!d\epsilon \nonumber \\
   && \times\,\frac{1}{\epsilon - h_{D}(\infty) + i \Lambda +
   \Delta\epsilon^{\alpha}(\infty)}\,\Lambda^{\alpha}, \label{pwblss}
\end{eqnarray}
If the external applied voltage assumes a step-like form, for
instance, $\Delta V^{\alpha}(t) = -\dea(t) = \Delta V^{\alpha}(1 -
\mbox{e}^{-t / a})$ with $a\rightarrow 0^+$, and $h_D(t)$ is not
affected by the fluctuation of $\sigma_D(t)$, Eq.~(\ref{paplus3})
would recover exactly Eq.~(\ref{paplus1}). In other cases,
Eq.~(\ref{paplus3}) provides an accurate and efficient approximation
for Eq.~(\ref{paplus1}), so long as $\Delta V^{\alpha}(t)$ do not
vary dramatically in time. Since the integration over energy in
Eq.~(\ref{paplus3}) can be performed readily by transforming the
integrand into diagonal representation, Eq.~(\ref{paplus3}) is
evaluated much faster than Eq.~(\ref{paplus1}). Due to its
efficiency and accuracy, Eq.~(\ref{paplus3}) is combined with
Eqs.~(\ref{qfinal0})-(\ref{p-alpha-minus}) to form the WBL
approximation for the dissipation functional $Q^{WBL}_{\alpha}$, and
thus recovers Eq.~(\ref{k-wbl-1}) of Sec.~\ref{schemes}.

As discussed in Sec.~\ref{schemes}, $Q^{WBL}_{\alpha}(t)$ depends
explicitly on $\Delta V^{\alpha}(t)$, $\sigma_D(t)$, $h_D(t)$ and
$\Lambda^{\alpha}$, where $h_D(t)$ is directly related to
$\rho_D(\mathbf{r},t)$ by the Poisson equation on $D$ subjected to
boundary conditions $\Delta V^{\alpha}(t)$, and $\Lambda^{\alpha}$
are associated with the DOS of $D$ near the surfaces $S_{\alpha}$.
Therefore in practice $Q^{WBL}_{\alpha}$ is a functional of
$\rho_D(\mathbf{r},t)$ only, \emph{i.e.},
\begin{eqnarray}\label{q-of-rho-d-1}
  Q^{WBL}_{\alpha}(t) &\equiv& Q^{WBL}_{\alpha}\Big[
  \sigma_D\!\left[\rho_D\right],
  \,h_D[\sigma_D\!\left[\rho_D\right],t], \nonumber \\
  && \Lambda^{\alpha}[\rho_D,t], \,\dva[\rho_D,t], \,t \Big].
\end{eqnarray}

\end{document}